# Virtual brightfield and fluorescence staining for Fourier ptychography via unsupervised deep learning

Ruihai Wang,[1,3,4] Pengming Song,[2,4,*] Shaowei Jiang,[3,4] Chenggang Yan,[1] Jiakai Zhu,[1,3] Chengfei Guo,[3] Zichao Bian,[3] Tianbo Wang,[3] and Guoan Zheng[2,3,*]

[1]*School of Automation, Hangzhou Dianzi University, Hangzhou 310018, China*
[2]*Department of Electrical and Computer Engineering, University of Connecticut, Storrs, CT, 06269, USA*
[3]*Department of Biomedical Engineering, University of Connecticut, Storrs, CT, 06269, USA*
[4]*These authors contributed equally to this work*
*\*Corresponding author: pengming.song@uconn.edu or guoan.zheng@uconn.edu*

**Fourier ptychographic microscopy (FPM) is a computational approach geared towards creating high-resolution and large field-of-view images without mechanical scanning. To acquire color images of histology slides, it often requires sequential acquisitions with red, green, and blue illuminations. The color reconstructions often suffer from coherent artifacts that are not presented in regular incoherent microscopy images. As a result, it remains a challenge to employ FPM for digital pathology applications, where resolution and color accuracy are of critical importance. Here we report a deep learning approach for performing unsupervised image-to-image translation of FPM reconstructions. A cycle-consistent adversarial network with multiscale structure similarity loss is trained to perform virtual brightfield and fluorescence staining of the recovered FPM images. In the training stage, we feed the network with two sets of unpaired images: 1) monochromatic FPM recovery, and 2) color or fluorescence images captured using a regular microscope. In the inference stage, the network takes the FPM input and outputs a virtually stained image with reduced coherent artifacts and improved image quality. We test the approach on various samples with different staining protocols. High-quality color and fluorescence reconstructions validate its effectiveness.**

The tradeoff between resolution and imaging field of view (FOV) is a major inconvenience for many microscopy applications. Fourier ptychographic microscopy (FPM) is a computational approach geared towards creating high-resolution and large FOV images without mechanical scanning [1]. In its original implementation, FPM illuminates the specimen from different incident angles and acquires the corresponding images using a low numerical aperture (NA) objective lens. With all low-resolution intensity acquisitions, a phase retrieval process can be used to synthesize the aperture information in the Fourier domain and generate a high-resolution complex image that retains the large FOV set by the low-NA objective. It also provides the ability to computationally correct optical aberrations post-measurement [2, 3].

One potential application for FPM is digital pathology, which acquires color whole slide images of stained histology slides for disease diagnosis. Current solution employs a robotic microscope for color image acquisition. In contrast, there are three advantages of using FPM for digital pathology. First, it can achieve high resolution and wide FOV without mechanical scanning. One can simply add an LED array to any existing microscope for setting up the platform. Second, in a conventional robotic microscope platform, autofocusing is needed to physically adjust the axial stage to bring the sample into focus when the sample is moving across different FOVs. If the captured image is out of focus, it would be challenging or impossible to bring it back to focus via post-acquisition processing. FPM addresses this issue by using a post-acquisition digital refocusing process, where a phase factor is introduced in the objective's pupil function to correct for the sample defocus [1, 4]. Third, the recovered phase information from FP further provides valuable information about the sample's local scattering and reduced scattering coefficients, which may benefit pathology diagnosis [5, 6].

To acquire color images of histology slides, FPM often requires sequential acquisitions with red, green, and blue illuminations. The color reconstructions often suffer from coherent artifacts that are generated, for example, by dust particles on the slide and lenses. The background of coherent images is, in general, not as clean as those captured with incoherent light. As a result, it remains a challenge to employ FPM for digital pathology applications, where resolution and color accuracy are of critical importance.

Here we report an unsupervised deep learning approach for 1) virtual color and fluorescence staining of FPM reconstructions, 2) reducing the

FPM coherent artifacts, 3) improving the overall image quality, and 4) shortening the acquisition time of color FPM by 67%. In our implementation, we employ a cycle-consistent adversarial network (cycleGAN) [7] with a multiscale structure similarity loss for unsupervised image to image translation. CycleGAN is a network that consists of two generator-discriminator pairs. Given an image in the source domain $A$, it can learn the conditional distribution of corresponding images in the target domain $B$, without seeing any pairs of corresponding images. CycleGAN has been applied in various imaging applications with great performance, including segmentation and classification [8], virtual staining [9-11], and blind deconvolution for fluorescence imaging [12].

takes the FPM input and creates a virtually stained image with reduced coherent artifacts and improved image quality. Different from the supervised approaches for virtual staining [13-15], paired examples are not needed in this implementation, i.e., no alignment or registration is needed for these two sets of images. In practice, it is often challenging to collect a large amount of paired training data. Unpaired training data, on the other hand, can be easily obtained at different instruments or from existing databases without tackling image registration issues.

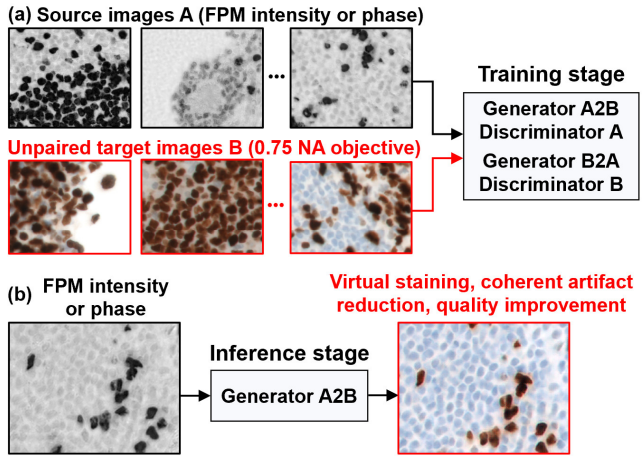

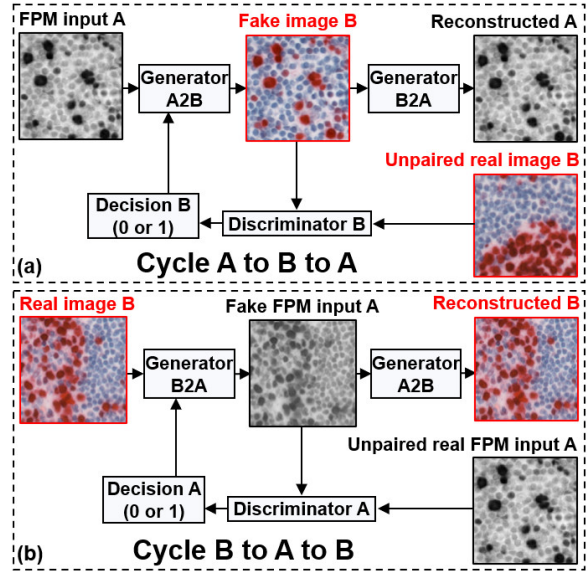

Fig. 1. The training and inference stages of the reported approach. (a) Two unpaired image sets are used to train two generator-discriminator pairs. (b) Generator A2B takes an FPM input and output a virtually stained image with reduced coherent artifacts and improved image quality.

Fig. 2. The cycleGAN structure for translating the input FPM intensity reconstruction $A$ into virtually stained incoherent brightfield image $B$.

Figure 1 shows the training and inference stages of our implementation. In the training stage (Fig. 1(a)), two sets of unpaired images are fed to the network to train two generator-discriminator pairs. The first set of images are FPM recovered intensity or phase, denoted as source images $A$. The second set of images are regular high-resolution microscopy images captured using a 20X, 0.75 NA objective, denoted as target images $B$. In the inference stage (Fig. 1(b)), one of the generators

Figure 2 shows the working principle of the network structure. For an FPM input image $A$, the generator $G_{AB}$ creates a fake incoherent microscopy image $B$ ($G_{AB}$: $A \rightarrow B$). Similarly, the generator $G_{BA}$ creates a fake FPM image $A$ based on a real incoherent microscopy image $B$ ($G_{BA}$: $B \rightarrow A$). Each generator has a corresponding discriminator, which attempts to tell apart its synthesized images from real ones, i.e., $D_A$ distinguishes $A$ from $G_{BA}(B)$, and $D_B$ distinguishes $B$ from $G_{AB}(A)$. As shown in Fig. 2, the key innovation of cycleGAN is to introduce cycle consistency constraint: $G_{BA}G_{AB}(A) \approx A$ and $G_{AB}G_{BA}(B) \approx B$ [7].

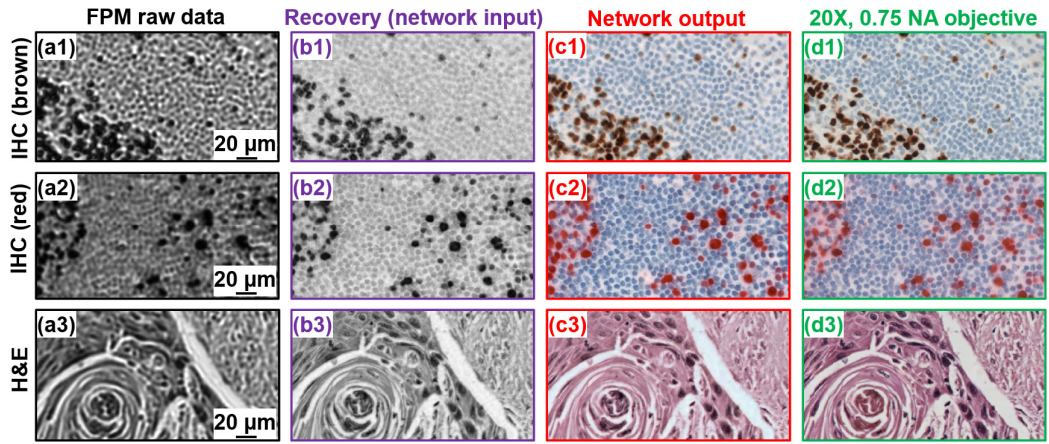

Fig. 3. Virtual staining for three different types of samples. (a) FPM raw data. (b) FPM recovery (network input). (c) Network output. (d) Ground truth.

We use the following loss function in our training process:

$$L(G_{AB}, G_{BA}, D_A, D_B) = L_{GAN-AB}(G_{AB}, D_A, A, B) + L_{GAN-BA}(G_{BA}, D_B, B, A)$$
$$+ L_{cyc}(G_{AB}, G_{BA},) + 0.1 \cdot (1 - msSSIM_g(G_{AB}(A), A))$$
$$+ 0.1 \cdot (1 - msSSIM_g(G_{BA}(B), B)), \quad (1)$$

where the first two terms $L_{GAN-AB}$ and $L_{GAN-BA}$ are the regular adversarial losses for the two generator-discriminator pairs. The third term $L_{cyc}$ is the cycle consistency loss that enforces $G_{BA}G_{AB}(A) \approx A$ and $G_{AB}G_{BA}(B) \approx B$. We adopt the same implementation of $L_{GAN-AB}$, $L_{GAN-BA}$, and $L_{cyc}$ as those in the Ref. [7]. The last two terms in Eq. (1) are used to reduce the structural changes between the input and output images. 'msSSIM$_g$' represents the multiscale structural similarity index (SSIM) of the green channel [16]. Adding these two terms can avoid color reversal and feature distortion between the input and output images [17].

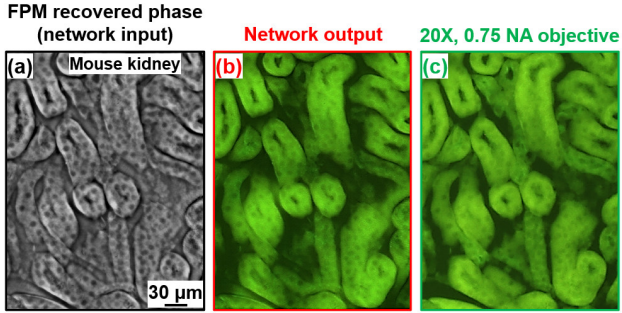

Fig. 4. Virtual fluorescence staining using the FPM recovered phase. (a) The FPM recovered phase. (b) The network output. (c) Ground truth.

In the first experiment, we test three types of samples for virtual FPM staining: 1) immunohistochemistry (IHC) brown stain, 2) IHC red stain, and 3) hematoxylin and eosin stain (H&E). For each sample type, we acquire 1500 high-resolution incoherent color images with 512 by 512 pixels each using a whole slide imaging system with a 20X, 0.75 NA objective lens [18]. With FPM, we use a setup with a 2X, 0.1 NA objective lens to acquire and recover the unpaired FPM images following the protocol in Ref. [3]. The recovered FPM intensity images are then segmented into 1500 small tiles with 512 by 512 pixels each. These two sets of unpaired images are fed into the deep neural network in Fig. 2 for the training process with 30 epochs. Figure 3 shows the brightfield staining results using new slides with the same staining protocol. The network outputs in Fig. 3 are similar to the ground-truth images captured using the 20X lens, validating its effectiveness.

Because of the structural similarity between the phase image and the fluorescence image, we have also tested the use of the FPM recovered phase to perform virtual fluorescence staining. In this experiment, we use the mouse kidney slides stained with Alexa Fluor 488 as our samples. Similar to the brightfield staining experiment, we acquire 1500 tiles of high-resolution fluorescence images using the whole slide imaging system. The FPM setup is used to generate ~1000 unpaired phase images. These sets of unpaired images are fed to the network. Figure 4(a) shows the input FPM phase and Fig. 4(b) shows the network output. Figure 4(c) shows the fluorescence ground-truth.

| SSIM | H&E | IHC (red) | IHC (brown) | Fluorescence |
|---|---|---|---|---|
| FPM and the ground-truth | 0.80±0.06 | 0.67±0.08 | 0.73±0.06 | 0.53±0.12 |
| Network output and the ground-truth | 0.88±0.04 | 0.81±0.06 | 0.86±0.04 | 0.66±0.08 |

Table 1. Image quality quantification between the FPM color / phase images and the network virtually stained output.

One challenge for adopting FPM for digital pathology is the coherent color artifacts. The reported approach can suppress the coherent artifacts and generate a color image similar to that captured by the regular incoherent microscope. In Fig. 5, we demonstrate the coherent artifact suppression and image quality improvement using the reported approach. Figure 5(a)-5(e) shows the FPM raw data, FPM intensity recovery, network output, FPM color image, and the ground truth, respectively. The line traces in Fig. 5(c1) and 5(c2) are shown in Fig. 5(f) and 5(g). We can see that the network output is in a good agreement with the ground truth intensity. The image quality of the network output is, in general, better than the FPM color image with sequential red, green, and blue illuminations.

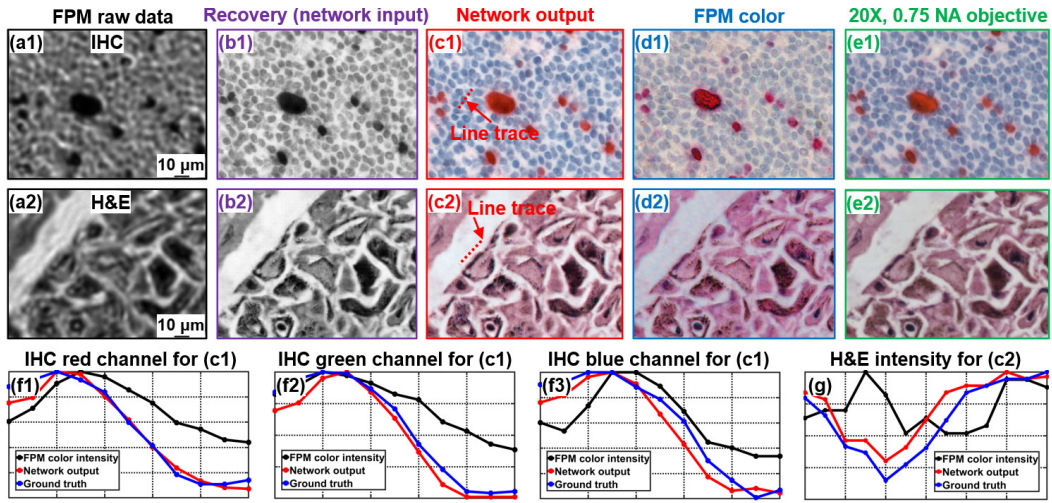

Fig. 5. Reducing color artifacts and improving image quality using the reported network. (a) FPM raw data. (b) FPM recovery (network input). (c) Network output. (d) FPM color image by sequential red, green, and blue LED illumination. (e) Ground truth.

We further quantify the image quality using the SSIM metric in Table 1. In the first three columns of Table 1, the SSIMs between the FPM color images and the ground-truths are listed in row 2 while the network outputs and the ground-truths are listed in row 3. In column 4 of Table 1, we use the fluorescence green channel as the ground truth and calculate the corresponding SSIMs. The listed results are averages of ~400 tiles for each type of specimen. We can see that the SSIMs of the network outputs are higher than those of FPM color and phase inputs.

One important advantage of FPM is to generate large FOV and high-resolution whole slide images without mechanical scanning. Such FPM recovered whole slide images can be fed into the reported network to generate virtually stained whole slide color images. Figure 6 shows the two whole slide color images generated by the reported network. The acquisition time is ~2 mins with a maximum synthetic NA of 0.55. We also compare the magnified view with the ground truth captured by the 20X, 0.75 NA objective lens.

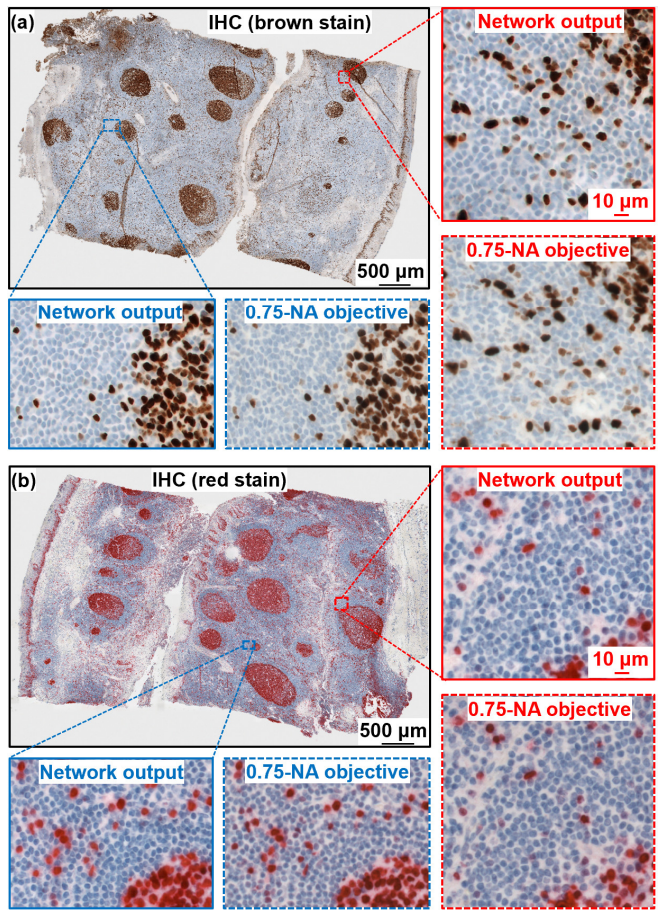

Fig. 6. Virtually stained FPM images of two types of IHC slides.

In summary, we have reported a data-driven approach for performing unsupervised image-to-image translation of FPM reconstructions. A multiscale SSIM loss term is added to the cycleGAN for reducing structural distortions between the input and output images. For brightfield color staining, the reported approach can shorten the acquisition time of FPM by 67%. Thanks to the structural similarity between the phase image and fluorescence images, we demonstrate virtual fluorescence staining using the FPM recovered phase as the input. Finally, the reported approach is not limited to translating the FPM images to those of incoherent modalities. It can be applied to other coherent imaging modalities, including coherent diffraction imaging, digital holography, real-space ptychography, where color images are often challenging to obtain. On-going efforts include testing this approach for ptychographic modulation microscopy [19-21].

**Funding.** This work is supported by the UConn Research Excellence Program and the UConn SPARK grant.

**Disclosures.** The authors declare no conflicts of interest.